# Spin-projection orientations in the plane square-lattice Ising model with periodic boundary conditions

You-gang Feng [*]

Department of Basic Science, College of Science, Guizhou University, Guiyang 550003,Cai Jia Guan, China

## Abstract

The periodic boundary conditions changed the plane square-lattice Ising model to the torus-lattice system which restricts the spin-projection orientations. Only two of the three important spin-projection orientations, parallel to the x-axis or to the y-axis, are suited to the torus-lattice system. The infinitesimal difference of the free-energies of the systems between the two systems mentioned above makes their critical temperatures infinitely close to each other, but their topological fundamental groups are distinct.

In 1944 Onsager obtained a solution of the critical temperature of the 2-dimensional Ising model with square-lattice system [1], since that time his solution has been regarded as an exact algebraic solution. A key point of the Onsager's solution is periodic boundary conditions. Under the conditions, as illustrated in FIG .1, the two lattices in sites $(-\infty, y)$ or $(x,-\infty)$ and $(+\infty, y)$ or $(x,+\infty)$ have the same spin-projection directions.

---

[*] E-mail: ygfeng45@yahoo.com.cn

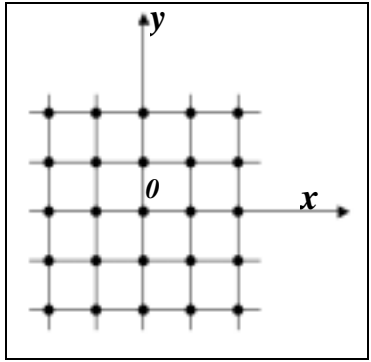

FIG. 1

The periodic boundary conditions changed the plane square-lattice system to a torus-lattice system which should be embedded in 3-dimensional Euclidean space. According to the scaling and self-similarity theory an ordered system is shrunk to one lattice point after a renormalization transform [2–3], but the torus-lattice system is not contractible because of its geometric topological structure [4]. FIG .2 shows that three simple closed curves on the torus only one's inner block of which, the curve B inner block, is a Kadanoff's block to become contractible [2 – 3]. But the singularity of the Onsager's equation indicated that the system certainly has the continuous phase transition which implied the system is ordered but contractible. However, there is a problem that the renormalization transform theory successfully treated the continuous phase transition of the plane square-lattice system and proved the system was contractible, which contradicts Onsager's torus-lattice system. Even though all they proved the system can become ordered, the difference is notable between the two methods. We noticed that there are only two spin states for the Ising model: spin-up and spin-down, but the important spin-projection orientations have three selections: it is normal to the plane, parallel to the x-axis or to the y-axis as illustrated in FIG.1. In fact, the partition function of the system is not affected by the projecting orientations because there is not any spin-projection orientation term in it. Even if it is in the computer simulation (Monte Carlo techniques) the orientations are neglected often [5 – 6]. We found that if the spin-projection orientation is normal to the plane with the periodic boundary conditions, the plane lattice becomes the torus lattice and the torus-lattice system cannot change to ordere. The reason is that if the system were ordered, the total spin-projection orientation of the system should be normal to the torus, but its projecting orientation is uncertain because the normal orientations of the torus are different and divergent everywhere. In addition, it is important that there is not the total spin-projection orientation of the system on the torus because neither the curve A interior nor the curve C interior (see FIG. 2 ) can shrink to a lattice point.

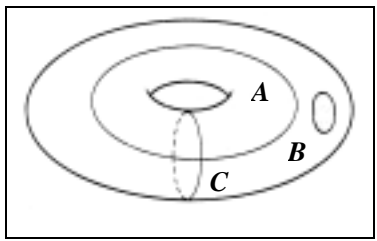

FIG.2

   However, if the spin-projection orientations are parallel to the x-axis or to the y-axis, the situation is completely different. Let the total spin of the system be $S$, the average lattice spin be $s$, and $s = S/N$, $N$ be the total number of the lattices in the system, $N \to \infty$, but $s \neq 0$, which accords with the thermodynamic limit condition. When the torus-lattice system becomes ordered, there is a continuous non-vanishing vector field on the torus, as illustrated in FIG 3 each little arrow represents an average lattice spin, this is a topological ordered state to which the singularity of the Onsager's equation corresponds. Obviously, such a system is unable to shrink, but it is ordered which shows that maybe such a kind of ordered system is not contractible. Nevertheless, a contractible system certainly will become ordered through the renormalization transform. An interesting thing is that the system without the periodic boundary conditions can be contractible through renormalization method. The periodic boundary conditions changed the geometric topological structure of the system [4], fortunately, the torus topological property allows the system to be ordered regarded as a topological ordered state, which second derivative of free-energy is still singular at the critical temperature contradicting to the topological phase transitions [7 – 8]. It means that the torus system's differential topological properties is equivalent to the plane system's differential topological properties, but their geometry topological properties are different.

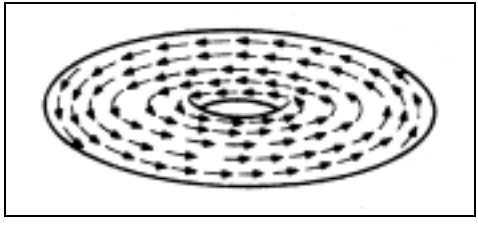

FIG . 3

In fact, the free-energy of the torus-lattice system is different from the free-energy of the plane square-lattice system. The partition function of the latter is defined as

$$Q = \sum_{s_i} \exp(-\frac{H}{k_B T}) = \sum_{s_i}^{(1)} \exp(-\frac{H}{k_B T}) + \sum_{s_i}^{(2)} \exp(-\frac{H}{k_B T}) \qquad (1)$$

$$H = -J \sum_{i,j} s_i s_j \qquad (2)$$

where $H$ is the system Hamiltonian, $J$ is a spin-coupling constant, $\sum_{i,j}$ denotes the sum over all possible nearest neighbor lattices, $\sum_{s_i}$ denotes the sum over all possible states of spins, $\sum_{s_i}^{(1)}$ denotes the sum over all possible states of spins with the periodic boundary conditions, $\sum_{s_i}^{(2)}$ denotes the sum over all possible states of spins failing to keep the conditions. The partition function of the first term in Eq.(1) is expressed by

$$Q_1 = \sum_{s_i}^{1} \exp(-\frac{H}{k_B T}) \qquad (3)$$

obviously, $Q_1$ is the partition function of the lattice system with the periodic boundary conditions, namely, of the torus-lattice system. The free-energy of the plane square-lattice system and the free-energy of the torus-lattice system are respectively given by

$$F = -k_B T \ln Q \qquad (4)$$

$$F_1 = -k_B T \ln Q_1 \qquad (5)$$

Because of the exponential function property, we have

$$Q \rangle 0, \qquad Q_1 \rangle 0, \qquad Q \rangle Q_1 \qquad (6)$$

the difference between $F_1$ and $F$ is

$$\Delta F = F_1 - F = k_B T \ln(\frac{Q}{Q_1}) \rangle 0 \qquad (7)$$

Under the thermodynamic limit condition $\Delta F$ changes to infinitesimal but zero, which makes the critical temperature of the torus-lattice system infinitely close to the critical temperature of the plane square-lattice system so that the Onsager's solution can be regarded as an exact solution of the plane square-lattice system. The infinitesimal difference $\Delta F \rangle 0$ shows that state, the little changing of the free-energy makes the topological structure transformation of the system so great that its fundamental group is completely different from the original[4].

In summary, the periodic boundary conditions changed the plane square-lattice Ising model to the torus-lattice system which possibility of continuous phase transition depends on the spin-projection orientations. Only two of the three important projecting orientations, parallel to the x-axis or to the y-axis, are suited to the torus-lattice system. The infinitesimal difference of free-energies of the systems between the two models mentioned above makes their critical temperatures infinitely close to each other, but their topological fundamental groups are distinct.